\documentclass[12pt]{article}
  \def\be{\begin{equation}}      
  \def\ee{\end{equation}}    \def\beq{\begin{eqnarray}}
      \def\eeq{\end{eqnarray}}

  \usepackage{graphicx}

  \newfont{\jnp}{cmss12 scaled\magstephalf}
  \newfont{\jig}{cmss10}
  \jnp

\begin{document}
{\Large Heavy Flavour Baryons in Hyper Central Model}\\

{\large Bhavin Patel$^*$, Ajay Kumar Rai $\ddag$ and P.C.Vinodkumar$^*$}\\

{$^*$Department of Physics, Sardar Patel University,
Vallabh Vidyanagar-388 120,\\ Gujarat, INDIA.\\
$\ddag$ Department of Applied Sciences and Humanities, Sardar
Vallabhbhai National \\ Institute of Technology, Surat-395 007, Gujarat,
INDIA} \\


PACS Nos \ {12.39.Jh\,-\,12.39.pn\,-\,14.20.kp}

\[Abstract\]\\
Heavy flavor baryons containing single and double charm
(beauty) quarks with light flavor combinations are studied using
the hyper central description of the three-body problem. The
confinement potential is assumed as hyper central coulomb plus
power potential with power index $\upsilon$. The ground state
masses of the heavy flavor, $J^P=\frac{1}{2}^+$ and
$\frac{3}{2}^+$ baryons are computed for different power index, $
\nu$  starting from 0.5 to 2.0. The predicted masses are found to
attain a saturated value in each case of quark combinations beyond
the power index $\nu=1.0$.\\

\section{Introduction}
Recent experimental observations of a family of doubly charms
baryons by SELEX, Fermi laboratory and most of the other charm
baryons discovered by CLEO experiments have created much interest
in the spectroscopy of heavy flavor baryons both experimentally
and theoretically
\cite{Giannini2001,Santopinto1998,Ebert2005,PDG2006,Athar2005,Eduarev2006,Avery1995,Bhavin2006,Gorelov2007}.
Baryons are not only the interesting systems to study the quark
dynamics and their properties, but also interesting from the point
of view of simple systems to study three body problems. Though
there are many theoretical attempts to study the baryons
\cite{Giannini2001,Santopinto1998,Ebert2005},
  many of them do not provide the form factors correctly that
 reproduces experimental data \cite{Giannini2001}. For this reason alternate
  schemes to describe the properties of baryons particularly in the heavy
  flavor sector are being attempted \cite{Giannini2001,Santopinto1998}.
Here, we employ the hyper central approach to study the three-body
problem, particularly the baryons constituting single and double
charm (beauty) quarks. The confinement
 potential is assumed in the hyper central co-ordinates of the coulomb plus
 power potential form. It should be mentioned that, hyper central potential
 contains the effects of the three body force. As suggested by lattice
 QCD calculations \cite{Gunnar2001} the three body forces  are important in
the study of baryons.
 For the low-lying resonance states it is good approximation
 to simply take the space wave functions of the hyper coulomb potential instead of
 seeking explicit numerical solution with hyperfine interaction.
\section{The Model}
A correct treatment for three body system is a long standing
problem in physics particularly in atomic and nuclear
  physics. Other three body systems of interest are the baryons containing three
  quarks. Typical interactions among the three quarks are studied
  using the two-body quark potentials such as the Igsur Karl Model,
  the Capstic and Isgur relativistic model, the Chiral model,
  the Harmonic Oscillator model etc. The three-body effects are incorporated in
  such models through two-body and three-body spin-orbit terms.
To describe the baryon as a bound state of three constituent
quarks, we define the configuration of three particles by two
Jacobi vectors $\vec{\rho}$ and
$\vec{\lambda}$ as \cite{Simonov1966}\\
\begin{equation}\label{}
    \vec{\rho}=\frac{1}{\sqrt{2}}(\vec{r}_1-\vec{r}_2)\,\,{;}\,\, \vec{\lambda}=\frac{1}{\sqrt{6}}(\vec{r}_1+\vec{r}_2-2\vec{r}_3)\\
\end{equation}
Such that
\begin{equation}\label{}
m_\rho=\frac{2\,\,m_1\,\, m_2}{m_1+m_2}
 \,\,{;}\,\,
m_\lambda=\frac{3\,\,m_3\,\,(m_1+m_2)}{2\,(m_1+m_2+m_3)}
\end{equation}\\
Here $m_1$, $m_2$ and $m_3$ are the constituent quark masses.
Further we introduce the hyper spherical coordinates which are
given by the angles
\begin{equation}\label{}
\Omega_\rho=(\theta_\rho,\phi_\rho)\,\,{;}\,\,
\Omega_\lambda=(\theta_\lambda,\phi_\lambda)
\end{equation}\label{}
together with the hyper radius, $x$ and hyper angle $\xi$ respectively defined by,\\
\begin{equation}\label{}
x=\sqrt{\rho^2+\lambda^2}\,\,{;}\,\,\xi=\arctan\left(\frac{\rho}{\lambda}\right)\\
\end{equation}\label{}
As a model Hamiltonian for baryons, we consider,
\begin{equation}\label{eq:404}
H=\frac{P^2_\rho}{2\,m_\rho}+\frac{P^2_\lambda}{2\,m_\lambda}+V(\rho,\lambda)=\frac{P^2}{2\,m}+V(x)\\
\end{equation}\label{}\\
Here the potential $V$ is not purely a two body interaction but it
contains three-body interactions also. If the interaction
potential is hyper central symmetric such that the potential
depends on the hyper radius \,$x$\,\,only, then the hyper radial
schrodinger equation corresponds to the hamiltonian given by
Eqn(\ref{eq:404}),  can be written as
\begin{equation}\label{eq:401}
\left[\frac{d^2}{dx^2}+\frac{5}{x}\frac{d}{dx}-\gamma(\gamma+4)\right]\phi_\gamma(x)=-2m[E-V(x)]\,\phi_\gamma(x)\\
\end{equation}\label{}\\where $\gamma$ is the grand angular quantum
number, m is the reduced mass \cite{Murthy1985} and it is defined
by
\begin{equation}\label{}
m=\frac{2\,\,m_\rho\,\, m_\lambda}{m_\rho+m_\lambda}
\end{equation}\label{}\\ and potential $V(x)$ is taken as\,\cite{Ajay2005}\\
\begin{equation}\label{}
V(x)=-\frac{\tau}{x}+\beta x^\nu+\kappa+V_{hyp}\left(x\right)\\
\end{equation}\label{}
Here the hyperfine part of the potential $V_{hyp}(x)$ is given by
\cite{Santopinto1998}
\begin{equation}\label{}
V_{hyp}(x)=A\,\,e^{-\alpha x}\sum\limits_{i{\neq}j}\sigma_i \cdot
\sigma_j
\end{equation}\label{}\\
where $\tau,\beta,A,\kappa$ and $\alpha$ are potential parameters.
The energy eigen value corresponding to Eqn(\ref{eq:401}), is
obtained using virial theorem for different choices of the
potential index $\nu$. The trial wave function is taken as the
hyper coulomb radial wave function given by \cite{Santopinto1998}
\begin{equation}\label{}
\psi_{\omega\gamma}=\left[\frac{(\omega-\gamma)!(2g)^6}{(2\omega+5)(\omega+\gamma+4)!}\right]^\frac{1}{2}(2gx)^\gamma
e^{-gx}
\end{equation}\label{}\\
The baryon mass in this hypercentral model is given by
\begin{equation}\label{}
M_B=\sum_{i=1}^{3} m_{i}+\langle H \rangle
\end{equation}\label{}\\
The constituent quark mass parameters employed in our calculations
are listed in Table-\ref{tab:01} along with other potential
parameters. Here $\kappa$ is found to be proportional to the
reduced mass, the flavor-color degree of freedom$(N_{f}N_{c})$ as
well as the strong coupling constant $\alpha_{s}$ as
\begin{equation}\label{}
\kappa\propto N_{c}N_{f} m \alpha_{s}(1+O(\alpha_{s}^2))
\end{equation}\label{}
It is found that the proportionality constant is equal to $0.41$
for the $qqQ$ systems and $0.32$ for the $QQq$ systems. The
computations are repeated for different choices of $\nu$, from 0.5
to 2.0 and the hyperfine interaction energy is treated
perturbatively.

\begin{figure}
\begin{center}\label{fig:01}
\includegraphics{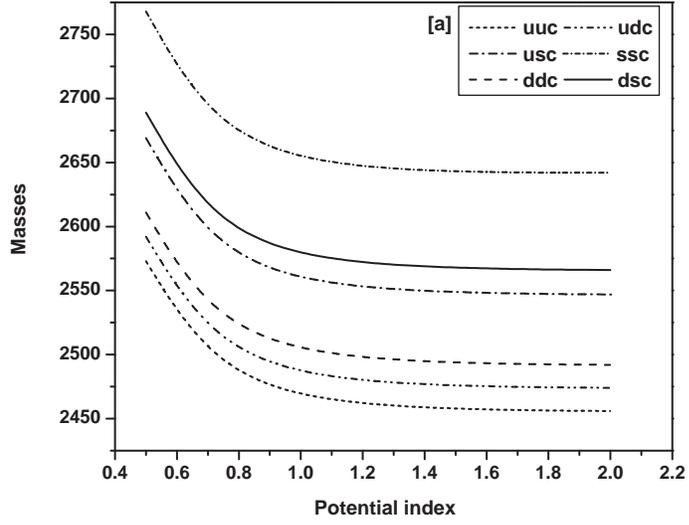}
\includegraphics{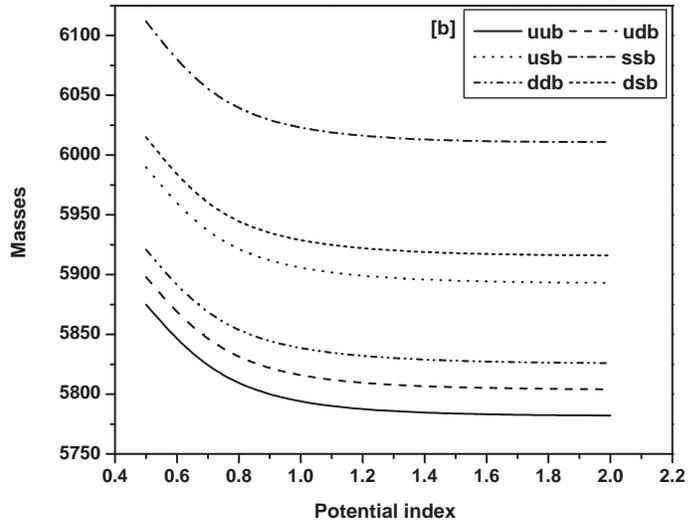}
\caption{Variation of spin average masses with potential index $\nu$ for single heavy baryons. [a] Single charm baryons, [b] Single beauty baryons.} \vspace{0.2cm}
\end{center}
\end{figure}

\begin{table} \caption{Quark Model Parameters} \vspace{0.01in}
\label{tab:01}
\begin{tabular}{lll}
\hline
&Quark masses & $m_{u}= 338$ \,(Me$V$) \\
              && $m_{d}= 350$ \,(Me$V$)\\
              && $m_{s}= 400$ \,(Me$V$)\\
              && $m_{c}=1394$ \,(Me$V$)\\
              && $m_{b}=4510$ \,(Me$V$)\\
 \hline
&Model parameter&$b=13.6$,  $\frac{\beta}{m\,\tau}=1 (MeV)^\nu$ \\
\hline
 &Spin-Spin interaction parameters& A$=140.7\,$ (Me$V$)\\
                                 &&$\alpha=850\,$(Me$V$)

\end{tabular}
\end{table}

\begin{figure}
\begin{center} \label{fig:02}
\includegraphics{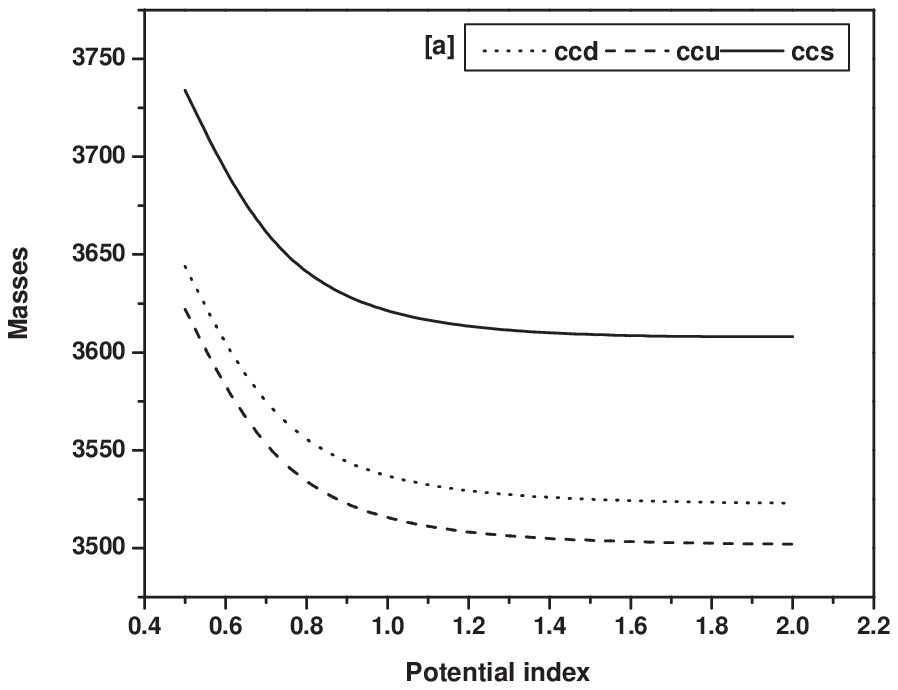}
\includegraphics{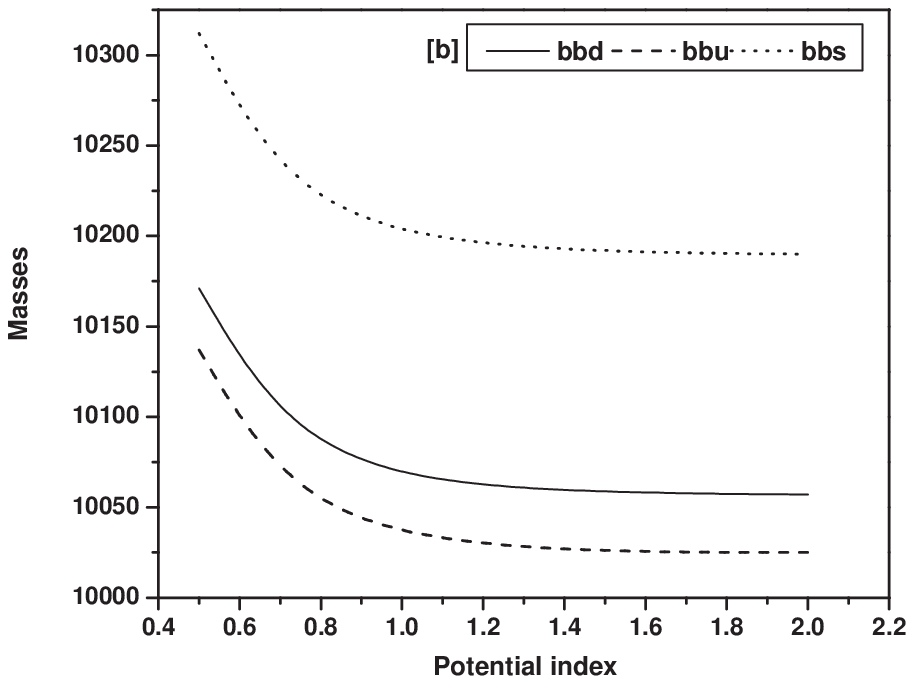}
\caption{Variation of spin average masses with potential index $\nu$ for doubly heavy baryons.[a] Doubly charm baryons, [b] Doubly beauty baryons. }
\vspace{0.2cm}
\end{center}
\end{figure}

\begin{table}
\begin{center} \caption{\it Single charm baryon masses(masses are in
MeV)} \vspace{0.001in} \label{tab:02}
\begin{tabular}{llllll}
 Baryon &P.I.($\nu$)
&{$\textbf{J}^P=\frac{1}{2}^+$}&Others&{$\textbf{J}^P=\frac{3}{2}^+$}&Others\\
\hline \hline
$\Sigma^{++}_{c}$&0.5&2539&2453\cite{Amand2006}&2607&$-$\\
(uuc)&0.7&2463&2454$\pm$0.18\cite{PDG2006}&2527&2518$\pm$0.6\cite{PDG2006} \\
&1.0&2432&2460$\pm$80\cite{Bowler1998}&2495&2440$\pm$70\cite{Bowler1998} \\
&1.5&2425&&2488& \\
&2.0&2425&&2488& \\
\hline
 $\Sigma^{+}_{c} $&0.5&2557&2451\cite{Amand2006}& 2627&$-$\\
(udc)&0.7&2480&2439\cite{Ebert2002}& 2546&2518\cite{Ebert2002}\\
&1.0&2449&2453\cite{Roncaglia1995}&2514&2520\cite{Roncaglia1995}\\
&1.5&2442&2452\cite{Mathur2002}&2507&2538\cite{Mathur2002}\\
&2.0&2442&2448\cite{Garcilazo2007}&2507&2505\cite{Garcilazo2007}\\
&&&$2453\pm0.4$\cite{PDG2006}&&$2518\pm2.3$\cite{PDG2006}\\
\hline
 $\Sigma^{0}_{c} $&0.5&2575&2452\cite{Amand2006}&2647
&$-$\\
(ddc)&0.7&2497&2454$\pm$0.18\cite{PDG2006}& 2566&2518$\pm$0.5\cite{PDG2006}\\
&1.0& 2466&& 2533&\\
&1.5& 2460&&2526 &\\
&2.0& 2460&& 2526&\\
\hline
 $\Xi^{+}_{c} $&0.5&2630&2466\cite{Amand2006}& 2708&$-$\\
(usc)&0.7&2550&2481\cite{Ebert2002}& 2625&2654\cite{Ebert2002}\\
&1.0&2518&2468\cite{Roncaglia1995}& 2591&2650\cite{Roncaglia1995}\\
&1.5&2512&2473\cite{Mathur2002}&2584&2680\cite{Mathur2002}\\
&2.0&2512&2496\cite{Garcilazo2007}&2584&2633\cite{Garcilazo2007}\\
&&&2468$\pm$0.4\cite{PDG2006}&&2647$\pm$1.4\cite{PDG2006}\\
&&&2410$\pm$50\cite{Bowler1998}&&2550$\pm$80\cite{Bowler1998}\\
\hline
$\Xi^{0}_{c}$ &0.5&2648&$2472$\cite{Amand2006} &2729&$-$\\
(dsc)&0.7&2567 &2471$\pm$0.4\cite{PDG2006}&2645&2646$\pm$1.2\cite{PDG2006}\\
&1.0&2536&&2611&\\
&1.5&2529&&2604&\\
&2.0&2529&&2604&\\
\hline $\Omega^{0}_{c}
$&0.5&2723&2698\cite{Amand2006}& 2813&$-$\\
(ssc)&0.7&2639&2698\cite{Ebert2002}& 2726&2768\cite{Ebert2002}\\
&1.0& 2607&2710\cite{Roncaglia1995}&2692&2770\cite{Roncaglia1995}\\
&1.5&2601&2678\cite{Mathur2002}&2684&2752\cite{Mathur2002}\\
&2.0&2601&2701\cite{Garcilazo2007}&2684&2759\cite{Garcilazo2007}\\
&&&2680$\pm$70\cite{Bowler1998}&&2660$\pm$80\cite{Bowler1998}\\
&&&2698$\pm$2.6\cite{PDG2006}&&\\
\hline
\end{tabular}
\end{center}
\end{table}

\begin{table}
\begin{center} \caption{\it Single beauty baryon masses(masses are in
MeV)} \vspace{0.001in} \label{tab:03}
\begin{tabular}{llllll}
\hline
 Baryon &P.I.($\nu$)
&{$\textbf{J}^P=\frac{1}{2}^+$}&Others&{$\textbf{J}^P=\frac{3}{2}^+$}&Others\\
\hline \hline
$\Sigma^{+}_{b}$&0.5&5862&5820 {\cite{Amand2006}}&5889&\\
(uub)&0.7&5803 &5770$\pm$70 {\cite{Bowler1998}}& 5828& 5780$\pm$70 {\cite{Bowler1998}}\\
&1.0&5778&$5808^{+02}_{-2.3}\pm$1.7\cite{Gorelov2007}&5801&$5829^{+1.6}_{-1.8}\pm$1.7\cite{Gorelov2007} \\
&1.5&5772&&5793& \\
&2.0&5772&& 5793& \\
\hline
$\Sigma^{-}_{b}$&0.5&5908&5820 {\cite{Amand2006}}&5937&$-$\\
(ddb)&0.7&5849 &$5816^{+01}_{-01}\pm1.7$ \cite{Gorelov2007}&5875&$5837^{+2.1}_{-1.9}\pm$1.7\cite{Gorelov2007}\\
&1.0&5823&&5847&\\
&1.5& 5816&&5840&\\
&2.0&5816&&5840&\\
\hline
 $\Sigma^{0}_{b}$& 0.5&5884&5624 {\cite{Amand2006}}&5912&$-$\\
(udb)& 0.7&5825 &5805 {\cite{Ebert2005}}&5851 &5834 {\cite{Ebert2005}}\\
&1.0&5800&5820 {\cite{Roncaglia1995}}&5823&5850 {\cite{Roncaglia1995}}\\
&1.5&5793&5847 {\cite{Mathur2002}}&5816&5871 {\cite{Mathur2002}}\\
&2.0&5793&5789 {\cite{Garcilazo2007}}&5816&5844 {\cite{Garcilazo2007}}\\
\hline
 $\Xi^{0}_{b}$ &0.5&5974&5624 {\cite{Amand2006}}&6007&$-$\\
(usb)&0.7&5913&5805 {\cite{Ebert2005}}&5943&5963 {\cite{Ebert2005}}\\
&1.0&5887&5820 {\cite{Roncaglia1995}}&5915&5980 {\cite{Roncaglia1995}}\\
&1.5&5880&5847 {\cite{Mathur2002}}&5907&5959 {\cite{Mathur2002}}\\
&2.0&5880&5789 {\cite{Garcilazo2007}}&5907&5967 {\cite{Garcilazo2007}}\\
&&&5760$\pm$60 {\cite{Bowler1998}}&&5900$\pm$80 {\cite{Bowler1998}}\\
\hline
$\Xi^{-}_{b}$&0.5&5997&5800 \cite{Amand2006}& 6032&$-$\\
(dsb)&0.7& 5936&& 5967&\\
&1.0&5909&&5938&\\
&1.5& 5903&&5931&\\
&2.0&5903&&5931&\\
\hline
 $\Omega^{-}_{b}$ &0.5&6092&6040 {\cite{Amand2006}}&6132&$-$\\
(ssb)&0.7&6028 &6065 {\cite{Ebert2005}}& 6064&6088 {\cite{Ebert2005}}\\
&1.0&6001&6060 {\cite{Roncaglia1995}}&6035&6090 {\cite{Roncaglia1995}}\\
&1.5&5994&6040 {\cite{Mathur2002}}&6028&6060 {\cite{Mathur2002}}\\
&2.0&5994&6037 {\cite{Garcilazo2007}}&6028&6090 {\cite{Garcilazo2007}}\\
&&&5990$\pm$70 {\cite{Bowler1998}}&&6000$\pm$70 {\cite{Bowler1998}}\\
\hline
\end{tabular}
%\end{indented}
\end{center}
\end{table}

\begin{table}
\begin{center} \caption{\it Doubly heavy baryon masses (masses are in MeV)}
\vskip 0.01 in \label{tab:04}
\begin{tabular}{llllll}
\hline
 Baryon & P.I.($\nu$)
&{$\textbf{J}^P=\frac{1}{2}^+$}&Others&{$\textbf{J}^P=\frac{3}{2}^+$}&Others\\
\hline \hline $\Xi^{++}_{cc}$&0.5&3583&$3612^{+17}$\cite{Albertus2006}&3660&$3706^{+23}$\cite{Albertus2006}\\
(ccu)&0.7&3505&3620\cite{Ebert2002}& 3578&3727\cite{Ebert2002}\\
&1.0&3475&3480\cite{Kselev2002}&3545&3610\cite{Kselev2002}\\
&1.5&3468&3740\cite{Tong2000}&3537&3860\cite{Tong2000}\\
&2.0&3468&3478\cite{Gershtein2002}&3537&3610\cite{Gershtein2002}\\
&&&3541\cite{Mattson2002}&&\\
\hline
 $\Xi^{+}_{cc}$&0.5&3604&3605$\pm$23\cite{Lewis2001}&3684&3685$\pm$23\cite{Lewis2001}\\
(ccd)&0.7&3525&3620\cite{Ebert2002}&3601 &3727\cite{Ebert2002}\\
&1.0&3494&3480\cite{Kselev2002}&3567&3610\cite{Kselev2002}\\
&1.5&3487&3740\cite{Tong2000}&3560&3860\cite{Tong2000}\\
&2.0&3487&3478\cite{Gershtein2002}&3560&3610\cite{Gershtein2002}\\
&&&3443\cite{Mattson2002}&&3520\cite{Mattson2002}\\
\hline
 $\Omega^{+}_{cc} $&0.5&3687&$3702^{+41}$\cite{Albertus2006}& 3782&$3783^{+22}$\cite{Albertus2006}\\
(ccs)&0.7&3604 &3778\cite{Ebert2002}& 3693&3872\cite{Ebert2002}\\
&1.0&3572&3590\cite{Kselev2002}&3659&3690\cite{Kselev2002}\\
&1.5&3566&3760\cite{Tong2000}&3651&3900\cite{Tong2000}\\
&2.0&3566&3590\cite{Gershtein2002}&3651&3690\cite{Gershtein2002}\\
&&&3733$\pm09$\cite{Lewis2001}&&3801$\pm09$\cite{Lewis2001}\\
\hline
$\Xi^{0}_{bb}$&0.5&10105&$10197^{+10}_{-17}$\cite{Albertus2006}& 10170&$10236^{+09}_{-17}$ \cite{Albertus2006}\\
(bbu)&0.7&10032 &10202 \cite{Ebert2002}&10092&10237 \cite{Ebert2002}\\
&1.0&10004&10090 \cite{Kselev2002}&10060&10130 \cite{Kselev2002}\\
&1.5&9998 &10300 \cite{Tong2000}&10053&10340 \cite{Tong2000}\\
&2.0&9998&10093 \cite{Gershtein2002}&10053&10133 \cite{Gershtein2002}\\
& & &10314$\pm$47 \cite{AliKhan2000}&&10333$\pm$45\cite{AliKhan2000}\\
\hline
$\Xi^{-}_{bb} $&0.5&10137&$10197^{+10}_{-17}$\cite{Albertus2006}&10206&$10236^{+09}_{-17}$ \cite{Albertus2006}\\
(bbd)&0.7&10063&10202 \cite{Ebert2002}&10127&10237 \cite{Ebert2002}\\
&1.0&10034&10090 \cite{Kselev2002}&10095&10130 \cite{Kselev2002}\\
&1.5&10028&10300 \cite{Tong2000}&10087&10340 \cite{Tong2000}\\
&2.0&10028&10314$\pm$47 \cite{AliKhan2000}&10087&10333$\pm$45\cite{AliKhan2000}\\
\hline
$\Omega^{-}_{bb}$&0.5&10269&$10260^{+14}_{-34}$ \cite{Albertus2006}& 10355&$10297^{+05}_{-28}$ \cite{Albertus2006}\\
(bbs)&0.7&10190 &10359 \cite{Ebert2002}& 10270&10389 \cite{Ebert2002}\\
&1.0&10160&10180 \cite{Kselev2002}&10236&10200 \cite{Kselev2002}\\
&1.5&10154&10340 \cite{Tong2000}&10228&10380 \cite{Tong2000}\\
&2.0&10154&10180 \cite{Gershtein2002}&10228&10200 \cite{Gershtein2002}\\
&&&10365$\pm$40 \cite{AliKhan2000}&&10383$\pm$39\cite{AliKhan2000}\\
\hline
\end{tabular}\vskip 0.001 in
\end{center}
\end{table}
\section{Results and Discussion}
The behavior of the spin average mass of the baryons with the
potential index $\nu$ in the case of $qqQ$ and $qQQ$ systems are
shown in Fig.1.(a,b) and Fig.{\ref{fig:02}}(a,b) respectively. It
is found that the mass of the baryon decreases as $\nu$ increases
and attain a saturated value beyond $\nu=1$. It may be due to the
saturation of effective inter quark interaction within the baryon
at potential index $\nu>1.0$. The computed results for the ground
state mass of the single heavy and double heavy baryons are
presented in Tables \ref{tab:02} and \ref{tab:03} respectively. We
compare our masses at this saturation $(\nu>1.0)$ with other
theoretical and existing experimental values.\\
Our results are found to be in accordance with the known
experimental as well as with other theoretical predictions in the
case of single heavy baryons at the mass saturation. The variation
with the PDG average values are just around 1.0 percentage in the
case of single charm baryons and less than 1.0 percentage in the
case of single beauty baryons. Consistency has also been found in
the case of double heavy systems with the potential index
$\nu\geq1.0$ with other theoretical predictions. Our results at
the saturated value of the masses are very close ($<1.0 $
percentage difference) to the theoretical predictions of S.
S.Gershtain et. al (2000) \cite{Gershtein2002} and V.V.Kiselev et.
al (2002) \cite{Kselev2002}. However, the predictions of C.
Albertus et. al (2006) \cite{Albertus2006} are found to be nearer
to our predicted masses at $\nu=0.5$. The recent observations of
SELEX group \cite{Mattson2002} on double charmed baryonic state
$\Xi_{cc}^{+}$ and $\Xi_{cc}^{*+}$ are found to be very close to
our predicted values. Our predicted mass difference
M($\Xi_{cc}^{*+}$)$-$M($\Xi_{cc}^{+}$) of 73.3 $MeV$ extremely
close to the lattice QCD prediction of 76.6 $MeV$
\cite{Bowler1998}. New experimental results are expected to
provide the masses of many of the double heavy flavor charm and
beauty baryons.\\

 {\bf Acknowledgement:} One of the
author P. C. Vinodkumar acknowledge the financial support from
University Grant Commission, Government of India under a Major
Research Project F. 32-31/2006 (SR).

\end{document}